\documentclass[preprint,showpacs,amsmath,amssymb,aps,doublespace]{revtex4}
\usepackage[dvips]{graphicx}
\usepackage{setspace}
\usepackage{amsmath}
\usepackage{amsfonts}
\usepackage{amssymb,color}

\begin{document}

\title{An epidemic process mediated by a decaying diffusing signal}

\author{Fernando P. Faria}
\email[e-mail: ]{fernandopereirabh@gmail.com}
\affiliation{Department of Physics of the Federal University of Minas Gerais,\\
Belo Horizonte, Minas Gerais, Brazil}
\author{Ronald Dickman}
\email[e-mail: ]{dickman@fisica.ufmg.br}
\affiliation{Department of Physics of the Federal University of Minas Gerais,\\
Belo Horizonte, Minas Gerais, Brazil}
\affiliation{National Institute of Science and Technology for Complex Systems,\\
Caixa Postal 702, 30161-970 \\ Belo Horizonte, Minas Gerais,
Brazil}

\begin{abstract}
We study a stochastic epidemic model consisting of elements
(organisms in a community or cells in tissue) with
fixed positions, in which damage or disease is
transmitted by diffusing agents (``signals") emitted by infected individuals.
The signals decay as well as diffuse; since they are assumed to be produced in large numbers,
the signal concentration is treated deterministically.  The model, which includes
four cellular states (susceptible, transformed, depleted, and removed),
admits various interpretations: spread of an infection or infectious disease,
or of damage in a tissue in which
injured cells may themselves provoke further damage, and as a description
of the so-called radiation-induced bystander effect, in which the signals are
molecules capable of inducing cell damage and/or death in unirradiated cells.
The model exhibits a continuous phase transition between spreading and nonspreading
phases.
We formulate two mean-field theory (MFT) descriptions of the model,
one of which ignores correlations between the cellular state and the signal concentration,
and another that treats such correlations in an approximate manner.
Monte Carlo simulations of the spread of infection on the square lattice
yield values for the critical exponents and the fractal
dimension consistent with the
dynamic percolation universality class.

\end{abstract}



\maketitle

\section{\label{intro}Introduction}

In many epidemic-like processes disease spreads via an
agent emitted by the affected elements (cells or organisms) themselves \cite{Mollis}.
Epidemics have been modeled extensively using deterministic
reaction-diffusion equations \cite{murray}, and stochastic particle
systems \cite{grass1,cardy,Linder,Souza}.
In the simplest epidemic models, such as the susceptible-infected-removed (SIR) and
susceptible-infected-removed-susceptible (SIRS) processes \cite{Bartlett},
disease is transmitted by contact between infected and healthy
organisms, without explicit representation of a transmitting agent.
But since the latter may have a dynamics of its own, typically involving diffusion
and decay, it is of interest to include this agent explicitly
in a more complete description, particularly
when the spatial structure of the epidemic is analyzed.

A similar observation applies to a viral infection, and
to the spread of damage in tissue following irradiation.
In the latter case, initially
affected cells may become sources of a signal that damages nearby cells,
which were not harmed in the initial event.  These secondary cells
may then become additional sources of the harmful signal.  Such a scenario has been
proposed for the radiation-induced bystander
effect (RIBE) \cite{BSDM,Mothersill97,Mothersill98}.
While the direct result is damage to some or all
of the irradiated cells, the long-term effect is characterized by damage or death of
unirradiated cells or ``bystanders".  It is thought that
irradiated cells release a signal (possibly a cytokine) that diffuses through the medium,
causing damage in previously healthy, unirradiated cells \cite{BSDM}.
Thus signal molecules in RIBE play a role analogous to a disease agent in an epidemic.

In this work we introduce an epidemic model in which damaged or infected elements
briefly emit signals; the latter diffuse and decay at a certain rates.
Healthy organisms or cells may become infected or damaged due to the presence of the signal,
and so may become new sources.  We formulate the model on a discrete two-dimensional space, the
square lattice.

Epidemic models with spatial structure and short-range interactions have been studied intensively
in recent years \cite {bai,grass1,Souza,sat1,sat2,Tome-Ziff}, and applied to the spread
of disease in humans and plants \cite{kuu,sander,gonci}.
Here we focus on processes initiated in a single cell or organism, or in
a localized region.  Key questions are then (1) the rate of spreading, as reflected
in the growth in the number of affected individuals, and their spatial distribution, and (2)
whether spreading continues indefinitely, limited only by the size of the available
region, or stops before attaining a size comparable to the of the system.
In the infinite system-size limit, the two regimes are separated by a phase transition.
In the supercritical (spreading) phase, there is a nonzero probability to spread indefinitely, whereas
in the subcritical (nonspreading) phase the process dies out with probability one.

Phase transitions in stochastic
epidemic models with spatial structure have received considerable attention; an important
example is the general epidemic process (GEP) \cite{Mollis, grass1}.
The GEP is essentially a stochastic susceptible-infected-removed (SIR) model
with spatial structure.  Initially, all individuals are susceptible (S) except
for one or a few infecteds (I).  Susceptibles with one or more I neighbors
become infected at a certain rate $\lambda$, while infecteds recover at rate $\mu$, after
which they are forever immune, hence removed (R) from interactions with other individuals.
Transmission (S+I $\to$ 2I) is typically restricted
to nearest-neighbor S-I pairs on a regular lattice or a network.
The GEP exhibits a phase transition as the ratio $\lambda/\mu$ is varied.
The supercritical phase is characterized by a growing active region, which invades regions
containing susceptibles and leaves behind an inactive region composed of individuals
in states S and R.  Activity is thus restricted to a ``ring" separating as yet unaffected and
formerly active regions.
(In a finite system, the final state is completely inactive.)
Analysis of the GEP shows that its critical behavior belongs to the dynamic percolation universality
class \cite{grass1,Tome-Ziff}.
If the process is modified so that a recovered individual can become
susceptible (SIRS model), it is possible to maintain an
active stationary state in which the processes of infection, recovery, and loss of
immunity occur continuously.  The SIRS model with spatial structure
exhibits a phase transition belonging to the directed percolation (DP) universality
class \cite{Henkel,Souza}, exemplified by the contact process \cite{harris,marro}.

In this work we study a model in which individuals may be in one of four states: susceptible (S),
transformed (T), depleted (D), and removed (R), the latter class designating individuals that
have died or are otherwise sequestered from the rest of the population.
The principal new feature  of our model is the mechanism
by which infection is transmitted: the transition from susceptible
to transformed is mediated by a signal released by cells in state T,
rather than via direct contact.  Such cells
may recover (becoming susceptible once again), may be removed, or may emit signals.  In the latter
case, the cell immediately enters state D, after which it
may either recover or be removed.  Although cells in states T or D may recover, there is
a finite probability of permanent removal.  Thus we expect that, as in the GEP,
the process will exhibit spreading and nonspreading phases, with activity
concentrated in a ring.  Assuming the phase transition is continuous, it is reasonable to expect the
critical behavior to be that of dynamic percolation.  Given the novel mode of transmission,
it is nonetheless of interest to verify this assumption.

The remainder of this paper is organized as follows.  We define the model in Sec. II, and
in Sec. III discuss two mean-field approaches, a simple one that ignores diffusion, and a
more detailed formulation that takes diffusion into account while still assuming spatial
homogeneity.
In Sec. IV we present simulation results for the phase diagram, critical behavior,
cluster properties, and spreading velocity.  A summary
and a discussion  of our results are provided in Sec. V.

\section{\label{model} Model}

The model is defined on a square lattice of $L^2$ sites, each of which hosts an
individual (an organism or a cell, depending on the choice of interpretation).
Individuals exist in one of four states: susceptible (S), transformed (T),
depleted (D), or removed (R).  In addition to the discrete state variable,
each site $(i,j)$ bears a signal concentration $C_{ij} \geq 0$.
Individuals emit signals upon making the transition from state T to state D; we adopt concentration
units such that each such event produces one unit of signal molecules.
The transitions between states are as follows: \\

(i) An individual in state S, at site $(i,j)$, has transition rates $\mu C_{ij}$ and
$\nu C_{ij}$ to states T and R, respectively.  These are the only
rates that depend on the signal concentration.\\

(ii) An individual in state T has transition rates  $w_{TS}$, $w_{TD}$ and $w_{TR}$, to states
S, D and R, respectively.\\

\begin{figure}[h]
\begin{center}
\includegraphics[height=9cm,width=12cm]{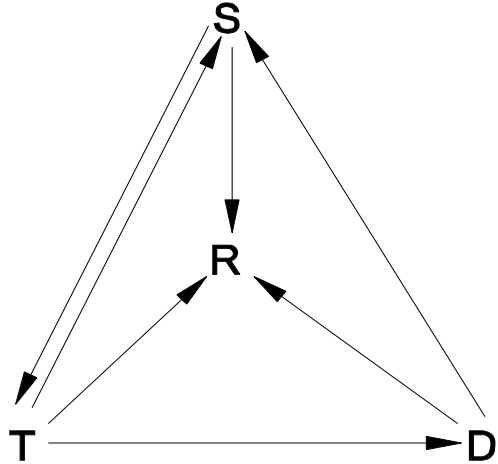}
\vspace{-2cm}

\caption{\footnotesize{States and allowed transitions.  The rates for transitions
out of state S are proportional to the local concentration of signal.  The latter is
produced when individuals move from state T to state D.}}
\label{states}
\end{center}
\end{figure}

(iii) An individual in state D has transition rates $w_{DS}$ and $w_{DR}$, to states S and R, respectively.\\
The states and transitions are summarized in Fig. \ref{states}; note that state R allows no escape.
\vspace{1em}

The signal concentration $C_{ij}(t)$ evolves via diffusion and decay.
We assume that the number of signal molecules is very large, so that
the concentration may be treated deterministically, via
the equation

\begin{equation}
\frac{d C_{ij}}{d t} = {\cal D}\Delta C_{ij} -\lambda C_{ij}
+ \sum_{k=1}^{n_{ij}} \delta (t-t_{k;ij}),
\label{dCijdt}
\end{equation}

\noindent where $\Delta$ denotes the discrete Laplacian,
$D$ is the diffusion rate, $\lambda$ is the decay
rate, $n_{ij}$ is the number of times site $(i,j)$ has made the
transition from T to D, and the $t_{k;ij}$ are the transition times.
Since the $n_{i,j}$ and the transition times are random variables, the $C_{i,j}$
are as well.

In the limit of very low signal concentration, the discrete nature
of the signal molecules makes an important contribution to the fluctuations.
Thus our continuum, deterministic description may not be suitable for the
low-concentration limit.  Another possibly troublesome aspect of the diffusion equation
is that, given a localized source at time zero, the concentration at any time $t>0$
is nonzero (albeit very small) at points arbitrarily far from the source.
While this is unphysical, we do not expect it to cause any significant effect
in the system under study.  Indeed, the diffusion equation is widely applied, with
apparent success, in modeling biological transport, and systems of reaction-diffusion
equations (including appropriate noise terms) have been found to yield
reliable predictions for critical properties of nonequilibrium systems \cite{Henkel}.

The model involves a rather larger set of parameters: the coefficients $\mu$ and $\nu$,
the diffusion and decay rates, and five additional transition rates.  It is nevertheless
clear that large values of $\mu$ and $w_{TD}/(w_{TR} + w_{TS})$, and small values of $\lambda$, favor
spreading.  Evidently, spreading can only occur
for ${\cal D}>0$.  Note that for $w_{DS} = 0$, there is no functional difference between states
D and R, and we have a simpler, three-state model.

We are primarily interested in an initial condition consisting of the origin in state D,
with an associated unitary signal concentration, and with all other sites in state S, and
free of any signal.  The ensuing evolution corresponds to an epidemic spreading from the
origin.
The current size of an epidemic can be defined as the
number of individuals in states other than S.  Of particular interest are the number $N_R(t)$
of removed individuals, the number $N_T(t)$ of individuals in state T, and the (spatial) average
signal concentration, $C(t)$.  The latter two quantities reflect the degree of spreading activity,
since, if both are zero, further advance of the epidemic is impossible.
In the spreading phase, starting from a
small, localized set of affected individuals, $N_R$ and $C$ grow without limit in an unbounded system,
whereas in the nonspreading phase these quantities cease to grow after a certain time.
In a finite system, $N_R$ and $C$, must eventually saturate, even in the spreading phase.
The distinction between spreading and nonspreading
phases is nonetheless evident in large, finite systems since,
in the spreading phase, a finite fraction of
individuals are eventually affected, whereas in the nonspreading phase the
final fraction of affected individuals is $\sim 1/L^2$  \cite{ker,daley}.
It is worth noting that, strictly speaking, an absorbing configuration corresponds to
one without transformed cells, and with the signal concentration everywhere zero.  Since $C$ decays at
a finite rate, such a situation can only obtain in the infinite-time limit.  The implications
for defining survival are discussed in Sec. IV.

For simplicity, we assume that the signal-dependent transitions (i.e., from S to either T or R)
have rates that are proportional to the local signal concentration.  Other dependencies are
conceivable; at the end of the next section we briefly consider rates proportional to $C^2$.

\section{\label{mf} Mean-field analysis}

In the simplest mean-field analysis we factorize the joint probability distribution for
the $N$-site system
into a product of single site probabilities,
and treat the signal concentration $C_{ij}$ as
independent of the state of the site.
Denoting the probabilities for site ($i,j$) to be in state S, T, D, or R by $S_{ij}$,
$T_{ij}$, $D_{ij}$, and $R_{ij}$ respectively, and the mean signal
concentration by $C_{ij}$, we then have:

\begin {eqnarray}
\frac{d C_{ij}}{dt} &=& {\cal D}\sum_{\langle kl,ij\rangle} (C_{kl}-C_{ij}) + w_{TD} T_{ij} - \lambda C_{ij}\\
\frac{d S_{ij}}{dt} &=& -(\mu + \nu) C_{ij} S_{ij}
+w_{TS} T_{ij} + w_{DS} D_{ij}\\
\frac{d T_{ij}}{dt} &=& \mu C_{ij} S_{ij} - (w_{TS}+ w_{TD} +w_{TR}) T_{ij}\\
\frac{d D_{ij}}{dt} &=& w_{TD}T_{ij} - (w_{DS}+w_{DR}) D_{ij}\\
\frac{d R_{ij}}{dt} &=& \nu C_{ij} S_{ij} + w_{TR} T_{ij} + w_{DR} D_{ij}
\label{mft}
\end{eqnarray}

\noindent where the sum in the first equation extends over the nearest neighbor
sites ($k,l$) of site ($i,j$).
If we take the continuum limit of these equations and let
${\bf X}({\bf r}) \equiv (C,S,T,D,R)$, we obtain a set of reaction-diffusion
equations, $\partial {\bf X}/{\partial t} = {\sf D} \nabla^2 {\bf X} + {\bf f}({\bf X})$, in which only
the element $D_{cc}$ of the diffusion matrix is nonzero.

We study a spatially uniform mean-field theory,
which corresponds to the fast-diffusion limit.  In this case the MF equations for the
site probabilities are as above (removing the subscripts $ij$ on all variables),
while the concentration satisfies

\begin{equation}
\frac{dC}{dt} = w T - \lambda C
\label{rhomfu}
\end{equation}

Given the large set of parameters, it is convenient to
fix all but one, which then plays the role of a control parameter.  Somewhat
arbitrarily, we choose $w \equiv w_{TD}$ (i.e., the rate at which transformed cells
emit signal and become depleted), as the control parameter, and denote its critical value by $w_c$.
We use the uniform analysis to set basic limits on survival of the spreading process,
by studying an epidemic scenario.
That is, we consider a very small initial source probability $D(0)$,
$S(0) = 1 - D(0) \simeq 1$,
and $T(0) = R(0) = 0$.  (We set $D(0) = C(0) = 10^{-13}$.)
Then, as $t \to \infty$, a non-vanishing value of $R$
corresponds to an epidemic in which a nonzero fraction of individuals are affected,
i.e., to the spreading phase.

\begin{figure}[h]
\begin{center}
\includegraphics[height=6.3cm,width=8.4cm]{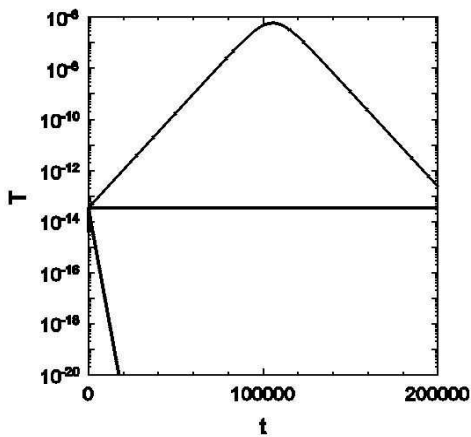}
\caption{\footnotesize{Transformed fraction $T(t)$ in
uniform mean-field theory.  Parameters $\lambda=0.05$, $\mu = 0.2$, $\nu = 0.1$,
$w_{TR} = w_{DR} = w_{TS} = w_{DS} = 0.2$, and (lower to upper) $w = 0.13$, $0.13333... = w_c$, and
$0.134$.}}
\label{f2new}
\end{center}
\end{figure}

\begin{figure}[h]
\begin{center}
\includegraphics[height=6.3cm,width=8.4cm]{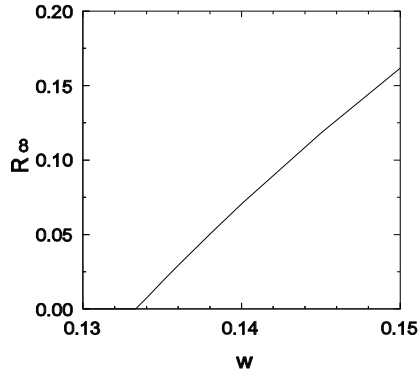}
\caption{\footnotesize{Uniform mean-field theory: limiting fraction of removed individuals
versus $w$.}}
\label{f3new}
\end{center}
\end{figure}

Figure \ref{f2new} shows the evolution of the transformed fraction $T(t)$;
in the nonspreading phase $T(t)$ decays monotonically, while in the spreading
phase it grows at intermediate times.  The depleted fraction $D(t)$
exhibits a similar behavior.  The fraction of removed individuals, $R(t)$, grows steadily
in the spreading regime, until it saturates; in the nonspreading regime it saturates at
a very small value (see Fig. \ref{f3new}).
In the spreading phase, the growth regime is followed by a crossover to exponential decay,
marking the end of the epidemic.  The crossover occurs due to the depletion
of susceptibles.

A simple analysis allows us to determine the
boundary between spreading and nonspreading phases.  Since we are interested in the
early stage of the evolution (following the initial transient) we set $S = 1$,
and seek a solution in which the probabilities $T$ and $D$, and the signal concentration
grow exponentially: $T(t) = T_1 e^{\gamma t}$, and similarly for $D$ and $C$.
Inserting these expressions in the MF equations yields

\begin{equation}
(\gamma + \lambda)(\gamma + w_T) = \mu w
\label{gamma}
\end{equation}
%\vspace{1em}

\noindent where $w_T = w + w_{TS} + w_{TR}$.
Equating $\gamma$ to zero yields the critical threshold:

\begin{equation}
w_c = \frac{\lambda (w_{TR} + w_{TS})}{\mu - \lambda}
\label{wcrit}
\end{equation}
\vspace{1em}

\noindent Note that spreading is impossible for $\mu < \lambda$, and that
$w_c$ is independent of parameters $\nu$, $w_{DS}$ and $w_{DR}$.
The above conclusions are verified
in numerical solutions of the (uniform) MF equations.

The original MF equations, Eq. (\ref{mft}), not only treat sites as independent, but
also treat the concentration at a site as independent of its state.  This is a rather
drastic approximation, because signal molecules are only created when a site undergoes
the transition T $\to$ D; other sites only acquire a nonzero signal concentration via
diffusion.  This approximation can be improved by introducing a concentration variable
$C_J$ for each site type; here $C_J$ denotes the mean concentration at a site, given that
it is in state $J$.  To derive a set of mean-field equations for the site probabilities
and the associated concentrations, we treat the amount of signal at a given site
as a discrete variable, $n$.  Let $P(J,n,t)$ denote the (joint) probability that a
site is in state J and has exactly $n$ units of signal.  Then the probability of state $J$ is
$P(J,t) = \sum_n P(J,n,t)$, and
$C_J(t) = \sum_n n P(J,n,t)/\sum_n P(J,n,t) = \sum_n n P(J,n,t)/P(J,t)$, so that

\begin{equation}
\frac{d C_J}{dt} = \frac{\sum_n n [dP(J,n,t)/dt]}{P(J,t)} - \frac{[dP(J,t)/dt] C_J(t)}{P(J,t)}
\label{dcjdt}
\end{equation}
\vspace{1em}

Consider, for example, state S.  There are contributions to $dP(S,n,t)/dt$ due to (1) decay of
the signal; (2) diffusion between the site and its neighbors; and (3) transitions between
state S and other states.  To treat diffusion at this level of approximation, we suppose that all
four neighbors of a given site have the same, average concentration,
$\overline{C(t)} = \sum_J P(J,t) C_J(t)$.  Then we have

\begin{eqnarray}
\frac{dP(S,n,t)}{dt} &=& \lambda [(n+1) P(S,n+1,t) - nP(S,n,t)]
\nonumber
\\
                       &+& 4D \overline{C}[P(S,n-1,t) - P(S,n,t)]
                       + 4D [(n+1) P(S,n+1,t) - nP(S,n,t)]
\nonumber
\\
                       &+& w_{TS} P(T,n,t) + w_{DS} P(D,n,t)
                       - n(\mu+\nu) P(S,n,t).
\end{eqnarray}

\noindent Summing over $n$, we find,

\begin{eqnarray}
\frac{\sum_n n [dP(S,n,t)/dt]}{P(S,t)} &=& -\lambda C_S + 4{\cal D}[\overline{C} - C_S]
- (\mu+\nu) \langle n^2 \rangle_S
\nonumber
\\
       &+& w_{TS} C_T \frac{P(T)}{P(S)} + w_{DS} C_D \frac{P(D)}{P(S)},
\label{dps1}
\end{eqnarray}

\noindent where $\langle n^2 \rangle_S \equiv \sum_n n^2 P(S,n.t)/P(S,t)$.
The second term in Eq. (\ref{dcjdt}) involves,

\begin{equation}
\frac{\sum_n [dP(S,n,t)/dt]}{P(S,t)} = -(\mu+\nu) C_S + w_{TS}\frac{P(T)}{P(S)} + w_{DS} C_D \frac{P(D)}{P(S)}.
\label{dps2}
\end{equation}
\vspace{1em}

\noindent Multiplying by $C_S$ and subtracting the result from Eq. (\ref{dps1}), we obtain

\begin{equation}
\frac{dC_S}{dt} = -\lambda C_S + 4{\cal D}[\overline{C} - C_S]
       + w_{TS} [C_T - C_S] \frac{P(T)}{P(S)} + w_{DS} [C_D - C_S] \frac{P(D)}{P(S)},
\label{dcsdt}
\end{equation}

\noindent where we have set var$(n) = \langle n^2 \rangle_S - C_s^2$ to zero, as is usual
in a mean-field approach.  Proceeding in the same manner, we find,

\begin{equation}
\frac{dC_T}{dt} = -\lambda C_T + 4{\cal D}[\overline{C} - C_T]
       + \mu C_S [C_S - C_T] \frac{P(S)}{P(T)},
\label{dctdt}
\end{equation}

\begin{equation}
\frac{dC_D}{dt} = -\lambda C_D + 4{\cal D}[\overline{C} - C_D]
       + w [1 + C_T - C_D] \frac{P(T)}{P(D)},
\label{dcddt}
\end{equation}

and

\begin{eqnarray}
\frac{dC_R}{dt} &=& -\lambda C_R + 4{\cal D}[\overline{C} - C_R]
       + \nu C_S [C_S - C_R] \frac{P(S)}{P(R)}
\nonumber
\\
       &+& w_{TR} [C_T - C_R] \frac{P(T)}{P(R)}
       + w_{DR} [C_D - C_R] \frac{P(D)}{P(R)}.
\label{dcrdt}
\end{eqnarray}

\begin{figure}[h]
\begin{center}
\includegraphics[height=8.8cm,width=11.8cm]{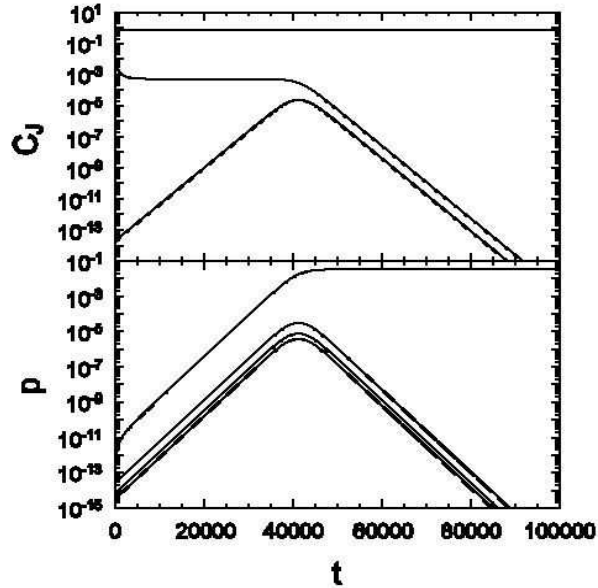}
\vspace{-1em}
\caption{\footnotesize{Diffusive mean-field theory: temporal evolution.
Lower panel (lower to upper): state probabilities $P(D,t)$, $P(T,t)$, total signal concentration
$C(t)$, and $P(R,t)$; upper panel (lower to upper): conditional concentrations $C_S$ and $C_T$
(indistinguishable at this scale), $C_R$, and $C_D$.
Parameters $\lambda=0.05$, $\mu = 0.2$, $\nu = 0.1$,
$w_{TR} = w_{DR} = w_{TS} = w_{DS} = 0.2$, ${\cal D} = 0.02$, and
$w = 0.2$, slightly greater than $w_c = 0.19587$. }}
\label{mfdfc}
\end{center}
\end{figure}

Numerical solution of this set of equations yields a critical threshold, $w_c$, which
decreases monotonically with diffusion rate ${\cal D}$, approaching the value of the
simple MF analysis as ${\cal D} \to \infty$.
Figure \ref{mfdfc} shows the evolution of the state probabilities and concentrations in a near-critical
system, as predicted by the diffusive mean-field theory (DMFT).
The predictions of
DMFT for $w_c$ are compared with simulation results in the following section.

\begin{figure}[h]
\begin{center}
\includegraphics[height=8.8cm,width=11.8cm]{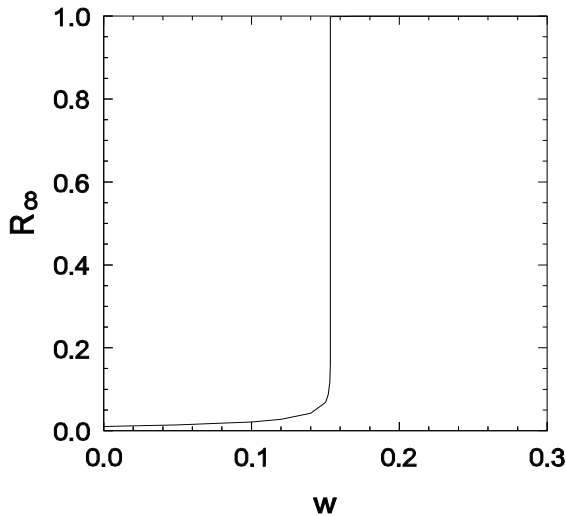}
\vspace{-1em}
\caption{\footnotesize{MFT: Final removal fraction $R_\infty$ versus $w$ for $\mu=10$, $C(0) = 0.01$,
when the transition rates from S to T and S to R are $\propto C^2$, with
other parameters as in Fig.~\ref{f2new}.  $R_\infty$ jumps from about 0.16 to 0.998 when $w=0.15335$.}}
\label{disc}
\end{center}
\end{figure}

An advantage of MFT is that it readily allows investigation of diverse scenarios.  We
use MFT to perform a preliminary study of a variant of the model defined above, in which
the rates for the transitions S $\to$ T and S $\to$ R are $\mu C^2$ and $\nu C^2$,
respectively, representing a situation in which healthy individuals are essentially
insensitive to very low signal concentrations.  In the epidemic context, this corresponds
to a situation in which small concentrations of a disease agent are effectively eliminated
by the immune system, whereas higher concentrations overwhelm it.
Experience with nonequilibrium phase
transitions in systems such as Schl\"ogl's second model \cite{schlogl}, leads one to
expect a discontinuous phase transition in this case.  This is indeed verified for
certain sets of parameters, featuring relatively large values of $\mu$; an example is shown in Fig.~\ref{disc}.
(Note that in this case the transition value $w_c$ depends on the initial signal concentration.)

\section{Simulations}

Simulations of the model defined in Sec. II are performed on square lattices of $L \times L$ sites.
These studies begin with all sites in state S (and having signal concentration zero) except for the
origin, which is in state D, with a signal concentration of unity.
Although the boundaries are open, the system size is chosen such that
sites on the boundary remain in state S, and have negligible signal
concentration, for the duration of the studies.
In the simulation algorithm,
at each step, corresponding to a time interval $\Delta t$, the
cellular states evolve as follows:\\

(i) if site $(i,j)$ is in state S, it remains in that state with probability
$p_s=\exp[-(\mu+\nu) C_{ij} \Delta t]$.  It makes a transition to state T
with probability $(1-p_s)\mu/(\mu+\nu)$, and to R with probability
$(1-p_s)\nu/(\mu+\nu)$.\\

(ii) if the site is in state T, it remains in this state with probability
$p_t=\exp[- w_T \Delta t]$.
The probability of a transition to state $J$ (=S, D, or R) is
$(1-p_t) w_{TJ}/w_T$.\\

(iii) if the site is in state D, it remains so with
probability $p_d=\exp[- w_D \Delta t]$. The
probability of a transition to state J (=S or R) is
$(1-p_d) w_{DJ}/w_D$.\\

(iv) sites in state R remain in this state.\\

\noindent In addition, the signal concentration is updated in accord with
Eq. (\ref{dCijdt}), that is, $C_{ij} \to C_{ij}' = \exp(-\lambda \Delta t) C_{ij}
+ {\cal D}\Delta t \sum_{kl}[C_{kl} - C_{ij}]$, with the additional contribution
$C_{ij} \to C_{ij} + 1$ at a step in which site $(i,j)$ makes a transition
from state T to state D.
Most of the studies reported below use $\Delta t = 0.4$.  We found that the results using
a smaller time step ($\Delta t = 0.2$) we essentially the same.  In studies of large ${\cal D}$
values, however, it was necessary to reduce the time step to 0.2 or 0.1 to maintain
reliability.

Since the
transition between spreading and nonspreading phases is apparently continuous,
we determine the critical point $w_c$ by searching for power-law behavior
of the level of activity $n(t)$,
the survival probability $P(t)$,
and the mean-square distance of removed cells cells from the origin, $R^2(t)$,
as functions of time.
The activity level $n(t)$ is conveniently defined as the number $N_T$ of sites in
state T, since growth of the active region requires transformed cells or a nonzero signal concentration.
We find that at long times, the behavior of the total signal concentration is similar
to that of $N_T$.

As noted in Sec. II, the definition of {\it survival} is not
completely obvious in the present model.  Our definition is based on the
continued increase in the number of R cells.  To begin, we study the distribution
of waiting times $t_w$ between successive events in which $N_R \to N_R + 1$.  Studying a
large number of realizations up to some maximum time, $t_M$,
we find that they
can be divided into two classes: those in which $N_R$ increases until the end, and those
in which this number saturates well before.  In the latter class, the final configuration
is devoid of T cells, and the total
signal concentration is extremely small, so that further growth in $N_R$ is
essentially impossible.
We find that in the class of realizations with sustained activity,
the probability distribution of waiting times, $P(t_w)$,
falls to zero above a certain value.  On this basis,
we define a maximum waiting time time, $t_{WM}$, somewhat larger than the value associated with
the cutoff of $P(t_w)$.
If, in a given realization, the waiting time $t_w$ exceeds $t_{WM}$, the system is taken to be
inactive, and the realization ends; the time of deactivation is taken
as that of the most recent transition to state R.
The survival probability $P(t)$
is then defined as the fraction of realizations that are active at time $t$.
(For the parameter set
analyzed below, we use $t_{WM} = 180$.)

The quantities $P(t)$, $n(t)$, and $R^2(t)$ are expected to follow \cite{grass1,grass2},

\begin{eqnarray}
n(t) & \sim & t^\theta ,\\
P(t) & \sim & t^{-\delta } ,\\
R^2(t) & \sim & t^{z_{sp}} ,
\end{eqnarray}

\noindent at the critical point, where $\theta $, $\delta $ and $z_{sp}$ are critical exponents associated
with spreading.
We expect the cluster generated by the critical process to be a fractal;
its radius of gyration should follow $R_g \sim n^{1/D_f}$, where
$D_f$ is the fractal dimension \cite{stauf}.

\begin{figure}[h]
\begin{center}
\includegraphics[height=8.0cm,width=11.8cm]{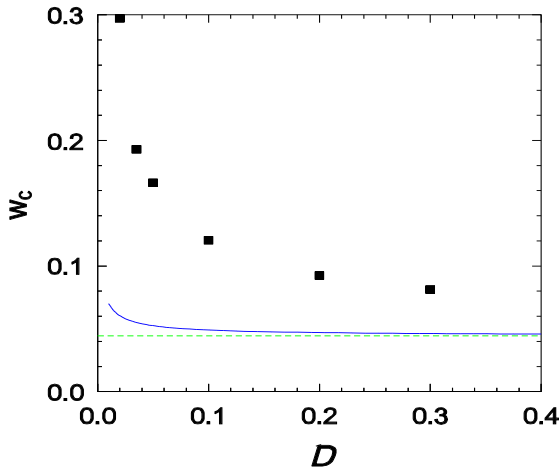}
\vspace{-2em}
\caption{\footnotesize{(Color online) Critical rate $w_c$ versus diffusion rate ${\cal D}$
for $\lambda=0.05$, $\mu = 0.5$, $\nu = 0.1$, and
$w_{TR} = w_{DR} = w_{TS} = w_{DS} = 0.2$.  Points: simulation;
solid curve: diffusive mean-field theory.  The dashed line indicates the
value predicted by simple MFT. Error bars are smaller than the symbols.
}}
\label{wcvda}
\end{center}
\end{figure}

\begin{figure}[h]
\begin{center}
\includegraphics[height=8.0cm,width=11.8cm]{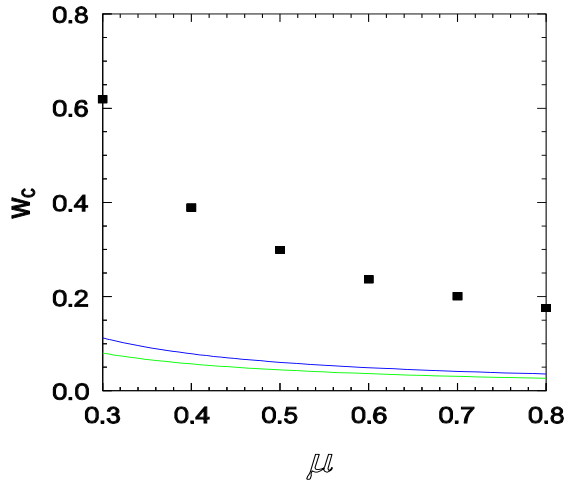}
\vspace{-2em}
\caption{\footnotesize{(Color online) Critical rate $w_c$ versus parameter $\mu$
for ${\cal D} = 0.02$ and other parameters as in Fig.~\ref{wcvda}.
Points: simulation;
solid curves: diffusive MFT (upper) and simple MFT (lower). Error bars are smaller than the symbols.
}}
\label{wcvmu}
\end{center}
\end{figure}

We perform detailed studies using the parameter values
$\lambda=0.05$, $\mu = 0.5$, $\nu = 0.1$, and
$w_{TR} = w_{DR} = w_{TS} = w_{DS} = 0.2$,
using system sizes $L$ of up to 600, and simulation times
of up to $t_M =10\,000$ time units.  (The number of
simulation steps is $t_M/\Delta t$.)
For each value of the diffusion rate ${\cal D}$ studied, we
determine the critical value $w_c$ using the power-law criterion for $P(t)$.
Specifically, we estimated $w_c$ using the local slope $\theta(t)$,
as described below.  The uncertainty in $w_c$, on the order of $5 \times 10^{-4}$,
reflects the range of $w$ values for which we cannot rule out an asymptotic
linear behavior of $\theta(t)$ versus $1/t$ at long times.
The resulting phase boundary is compared against mean-field theory
in Fig.~\ref{wcvda}.  Mean-field theory furnishes the correct order of
magnitude, but underestimates $w_c$, especially for small values of ${\cal D}$.
The diffusive MFT furnishes a slight improvement over simple MFT.
For ${\cal D} \to \infty$,
the mean-field prediction converges to 2/45 = 0.0444...; simulations in this limit
(using a spatially uniform signal concentration) yield $w_c = 0.045(1)$, consistent with MFT.
Figure \ref{wcvmu} shows a similar comparison for $w_c$ as a function of $\mu$, for ${\cal D} = 0.02$;
in this case the DMFT prediction is about a factor of five smaller than the simulation value.

We also determined $w_c$ for a rather different set of parameters:
$\nu=0.51$, $\mu = 0.79$, $\lambda = 0.1$, ${\cal D} = 2.94$, and
$w_{DR} = w_{DS} = w_{TR} = w_{TS} = 0.0079$.  In this case simulation yields
$w_c=3.15(5) \times 10^{-3}$, while simple and diffusive MFT yield $w_c = 2.29 \times 10^{-3}$ and
$2.31 \times 10^{-3}$, respectively.  The closer agreement between simulation and MFT in this case
can be attributed to the higher diffusion rate.

\subsection{\label{Res1} Critical behavior}

Following preliminary studies which indicated that $w_c \simeq 0.297$, (for the parameter set mentioned
above, with ${\cal D} = 0.02$), we performed a more detailed study using $L=360$ and $t_M = 5\,000$,
with 72\,000 realizations for each value of $w$ studied.
A large number of realizations is necessary to obtain a clear result for $w_c$ and the critical
exponents.  Using a set of 10$^3$ or 10$^4$ realizations (a number that would be sufficient for
studying the contact process, for example) the results are dominated by fluctuations.
We believe that this is due to the multi-step nature of propagation, and in particular, to diffusion.
For the relatively small diffusion rate used here, the initial stages of propagation depend on
rare events, whereas for large values of ${\cal D}$, we expect that long simulations of large
systems would be needed to observe the crossover from mean-field-like behavior to the asymptotic
scaling regime.

\begin{figure}[h]
\begin{center}
\includegraphics[height=6cm,width=7.5cm]{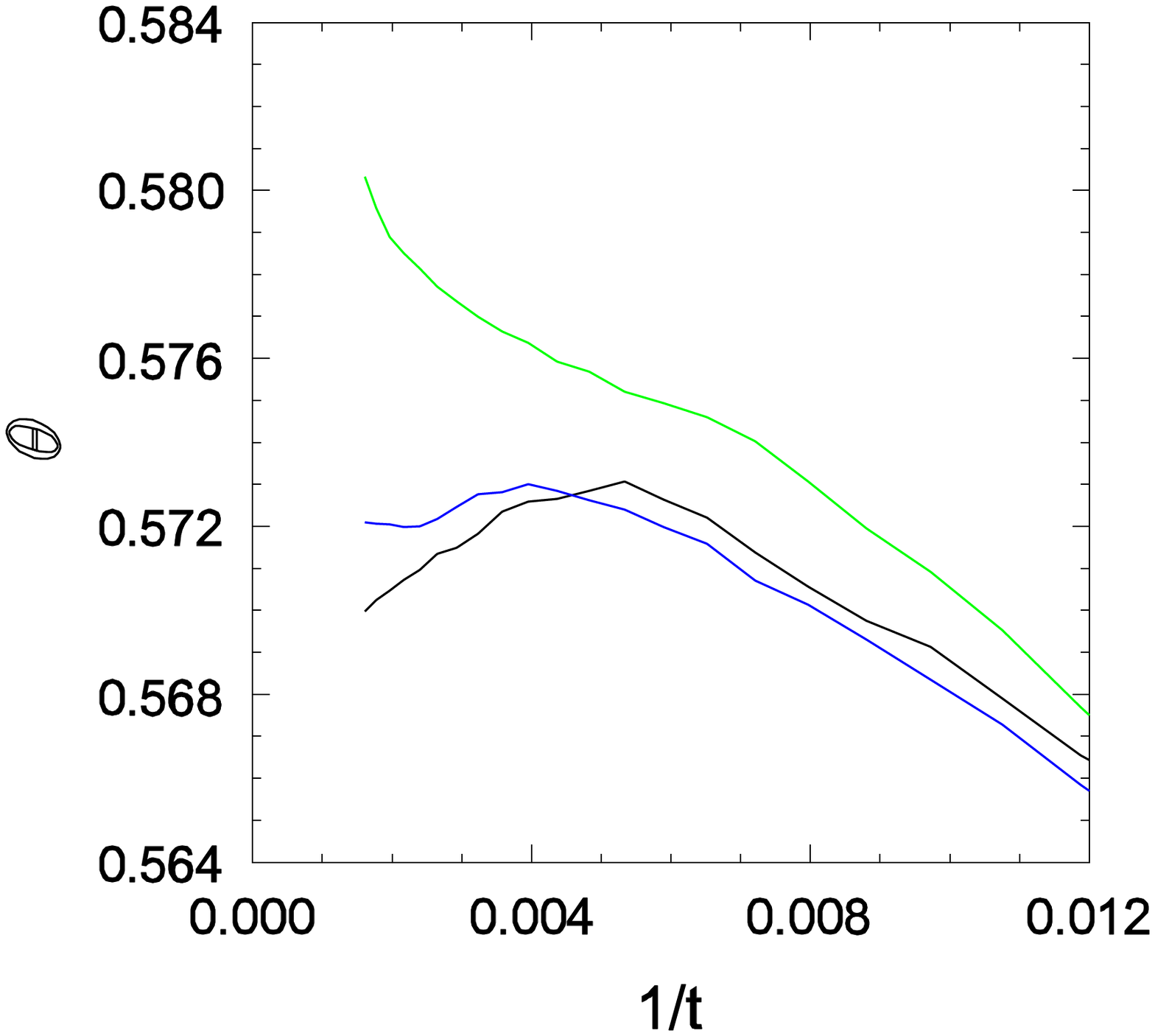}
\includegraphics[height=6cm,width=7.5cm]{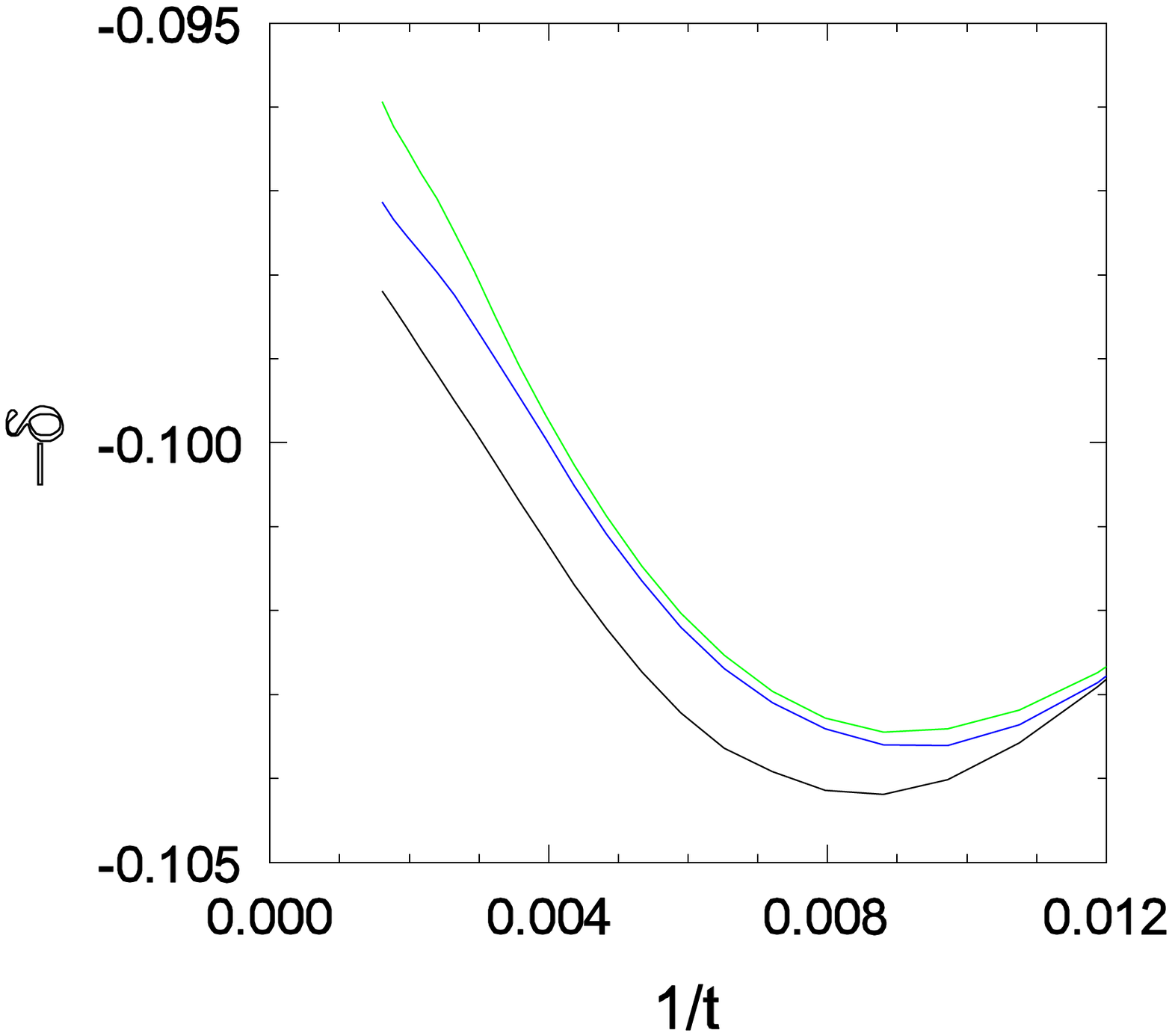}
\includegraphics[height=6cm,width=7.5cm]{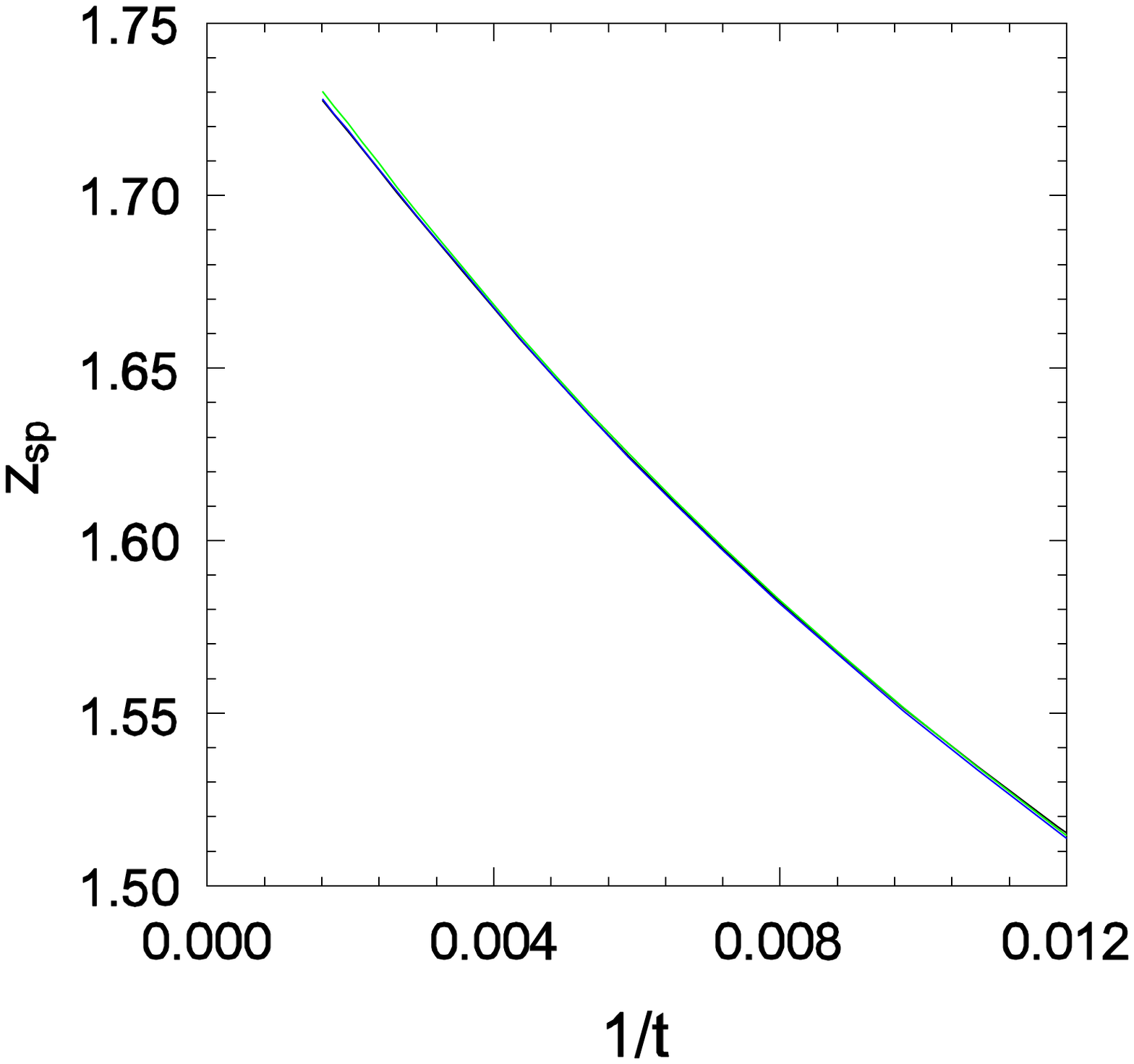}
\caption{\footnotesize{Simulation: local slopes $\theta(t)$, $\delta(t)$, and $z_{sp}(t)$ for
(lower to upper, on left) $w = 0.297$, 0.2972, and 0.2975.  Parameters: ${\cal D} = 0.02$,
$\lambda=0.05$, $\mu = 0.5$, $\nu = 0.1$, and
$w_{TR} = w_{DR} = w_{TS} = w_{DS} = 0.2$.
}}
\label{ls}
\end{center}
\end{figure}

The spreading exponents $\theta$, $\delta$ and $z_{sp}$
are estimated via analysis of the
local slopes, $\theta(t)$, etc.  For example, $\theta(t)$ is
defined as the inclination of a least-square linear fit to the data for $n(t)$ (on logarithmic scales),
on the interval $[t/a, \,at]$.  (The choice $a$ represents a compromise between
high resolution, for smaller $a$, and insensitivity to fluctuations, for larger values;
we use $a$ in the range 2-3.)
We estimate the exponents by plotting the local slopes
versus $1/t$ and extrapolating to $1/t= 0$.  Since a supercritical
process leads to local slopes that curve upward, and vice-versa,
we seek the value of $w$ associated with the least curvature.
The local slopes are plotted in Fig.~\ref{ls}.
On this basis of these results we estimate the critical point as $w_c=0.2972(1)$,
and find $\theta = 0.573(5)$, $\delta = 0.096(1)$ and
$z_{sp} = 1.768(2)$.  (The estimate for $w_c$ is based on the data for $\theta(t)$.)
The results for the exponents compare reasonably well with the literature values,
$\theta=0.586$, $\delta = 0.092$, and $z_{sp}=1.771$ for dynamic percolation
in two dimensions \cite{Hav,Munoz}.
The main source of uncertainty in the exponent estimates is the uncertainty in $w_c$
itself.
The exponents obey the scaling relation
of dynamic percolation, $\theta = z_{sp}-2\delta-1$, to within uncertainty.

\begin{figure}[h]
\begin{center}
\includegraphics[height=6.0cm,width=8.0cm]{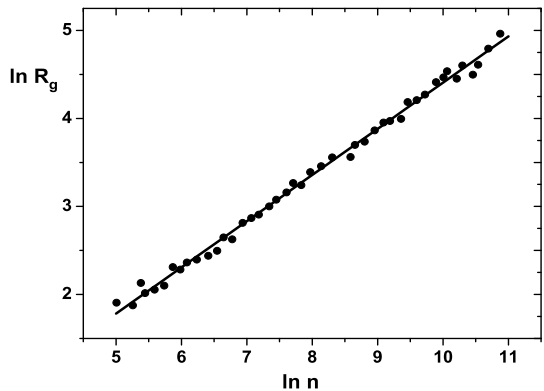}
\caption{\footnotesize{Simulation: radius of gyration $R_g$ versus cluster size $n$ at the critical
point for parameters as in Fig.~\ref{ls}, with $w=0.2972$.}}
\label{Fractal}
\end{center}
\end{figure}

To determine the fractal dimension of the critical cluster, we studied the radius of gyration
$R_g$ of the final cluster as a function of its size, $n$,
in a set of 500 realizations using $t_M = 5000$.
One expects that at the critical point,
$R_g \propto n^{1/D_f}$, for $n \gg 1$ \cite{stauf}.  A least-squares linear fit to the data
(see Fig.~\ref{Fractal}) yields $R_g \sim n^{0.525(6)}$, corresponding to a fractal dimension
of $D_F = 1.91(2)$.  The value for two-dimensional percolation is 91/48 $\simeq$ 1.896 \cite{stauf}.

\subsection{Subcritical regime}

In the subcritical (nonspreading) regime, three quantities of interest are the mean lifetime $t_p$, the mean
(final) cluster size $n$ and the mean radius of gyration $R_g$ of the final cluster.
One expects that, in the neighborhood the critical point, these quantities scale as \cite{Hav},

\begin{equation}
{t_p \propto \Delta^{-\nu_{\parallel }}},\\
\quad n \propto \Delta^{-\zeta }, \\
\quad R_s\propto \Delta^{-\nu_{\perp }}\\
\end{equation}

\noindent where $\Delta = w_c - w$, and $\zeta =\gamma\nu_{\parallel }/\nu_{\perp }$, with $\gamma$ the
percolation critical exponent governing the divergence of the mean cluster size.

\begin{figure}[h]
   \begin{minipage}[t]{0.5 \linewidth}
        {\includegraphics[width=\linewidth]{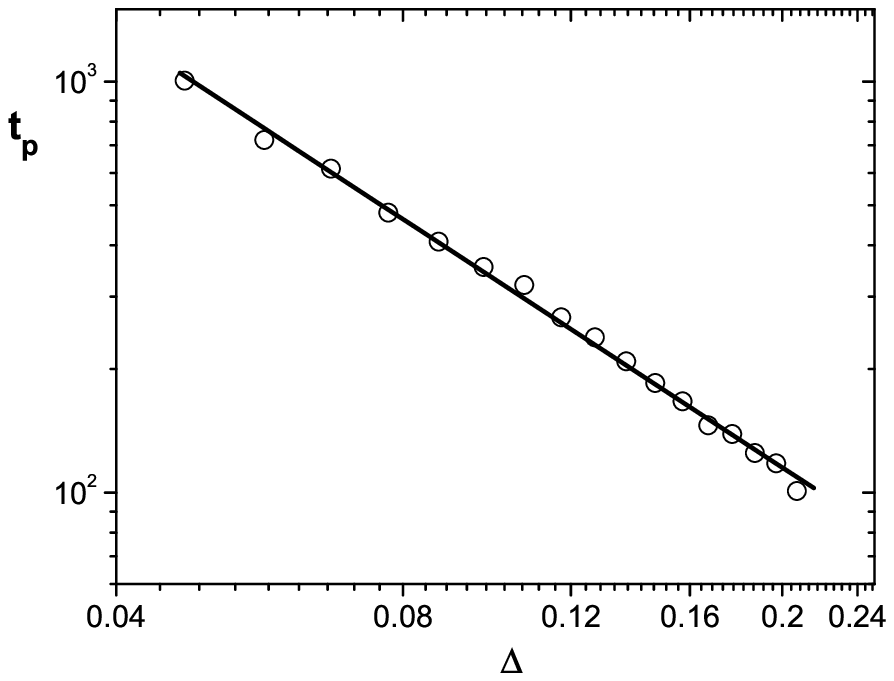}}
       \end{minipage}\hfill
       \begin{minipage}[t]{0.5 \linewidth}
         {\includegraphics[width=\linewidth]{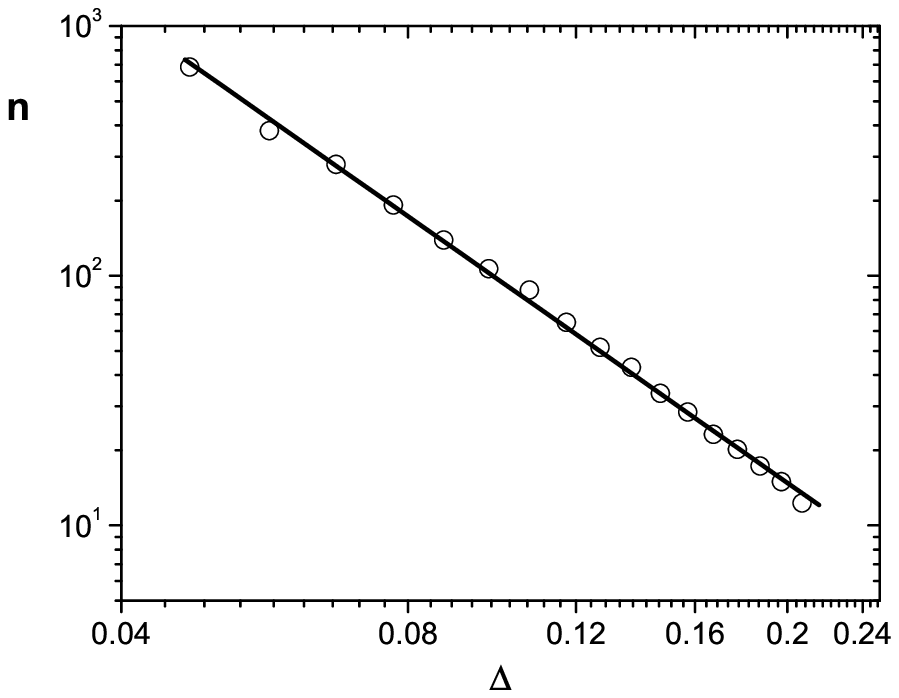}}
       \end{minipage}
        \begin{minipage}[t]{0.5 \linewidth}
         {\includegraphics[width=\linewidth]{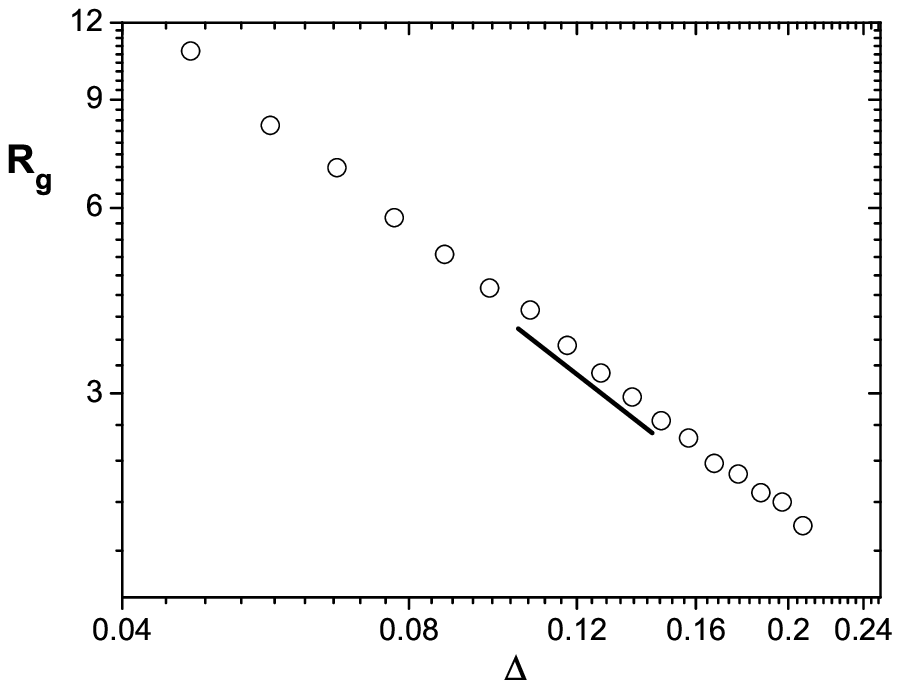}}
       \end{minipage}\hfill
       \caption{\footnotesize{Simulation: survival time $t_p$, mean cluster size
       $n$, and mean radius of gyration $R_g$ versus $\Delta = w_c - w$, in the subcritical
       regime, for parameters as in Fig.~\ref{ls}.  The slopes of the regression lines are
       1.52 ($t_p$), 2.68 ($n$), and 1.29 ($R_g$).}}
\label{crit}
\end{figure}

We estimate the exponents $\nu_{\parallel }$, $\zeta $ and $\nu_{\perp }$ using simulations
with system size $L=1000$, and 1000 realizations for each value of $w$ studied,
for the parameter set defined above
(see Fig. \ref{crit}).  The simulation results yield the estimates
$\nu_{\parallel }=1.52(2)$, $\nu_{\perp }=1.29(3)$, and $\zeta=2.69(3)$.
(We note, however, that the result for $\nu_\perp$ should not be taken
very seriously, given the small values of $R_g$.)
The reference values for dynamic percolation in two dimensions are
$\nu_{\parallel }=1.506$, $\nu_{\perp }=4/3$, and $\zeta=2.698$ \cite{Hav}.

\subsection{Clusters and spreading}

\begin{figure}[h]
\begin{center}
\includegraphics[height=6cm,width=7.5cm]{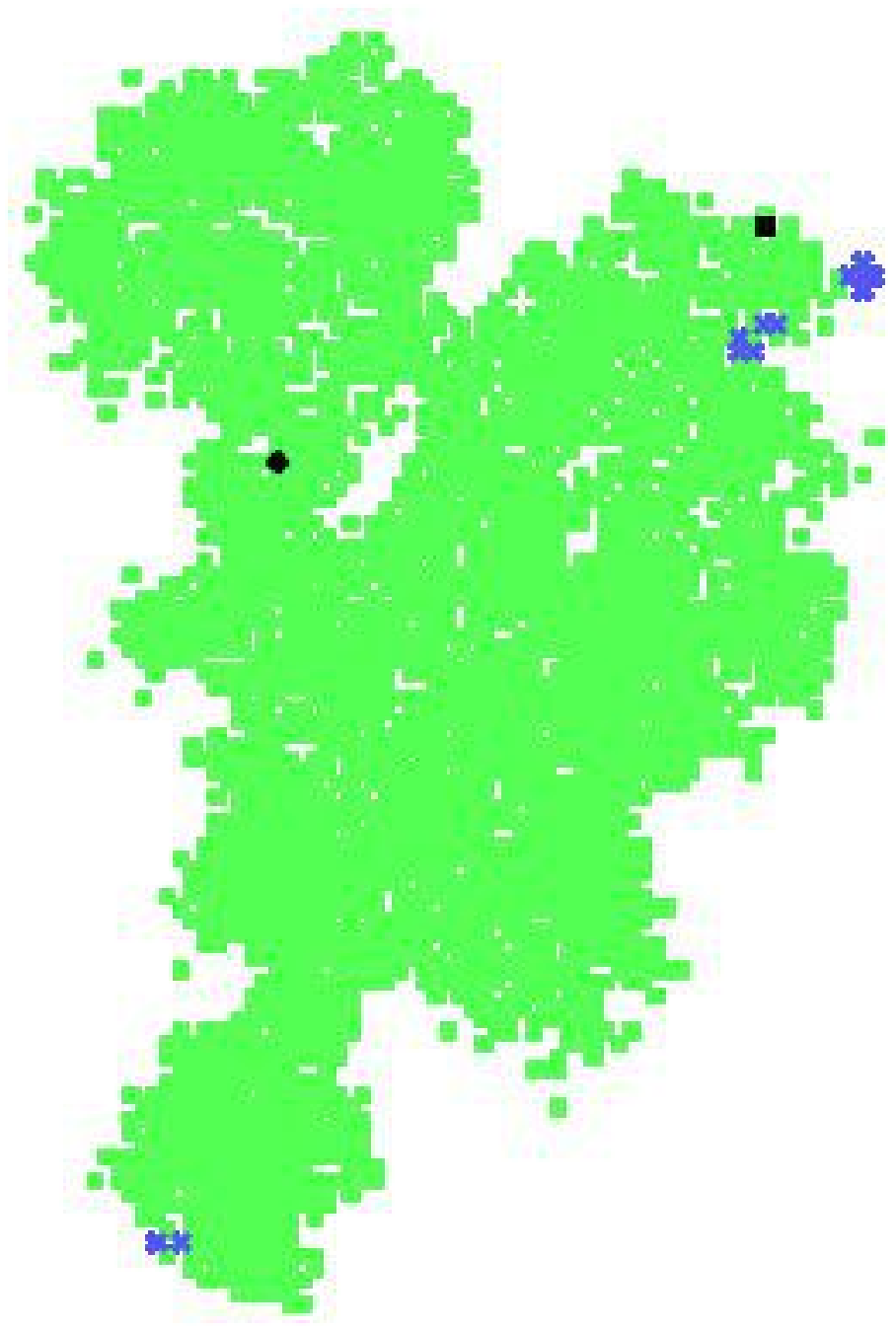}
%\vspace{-2cm}
\includegraphics[height=6cm,width=7.5cm]{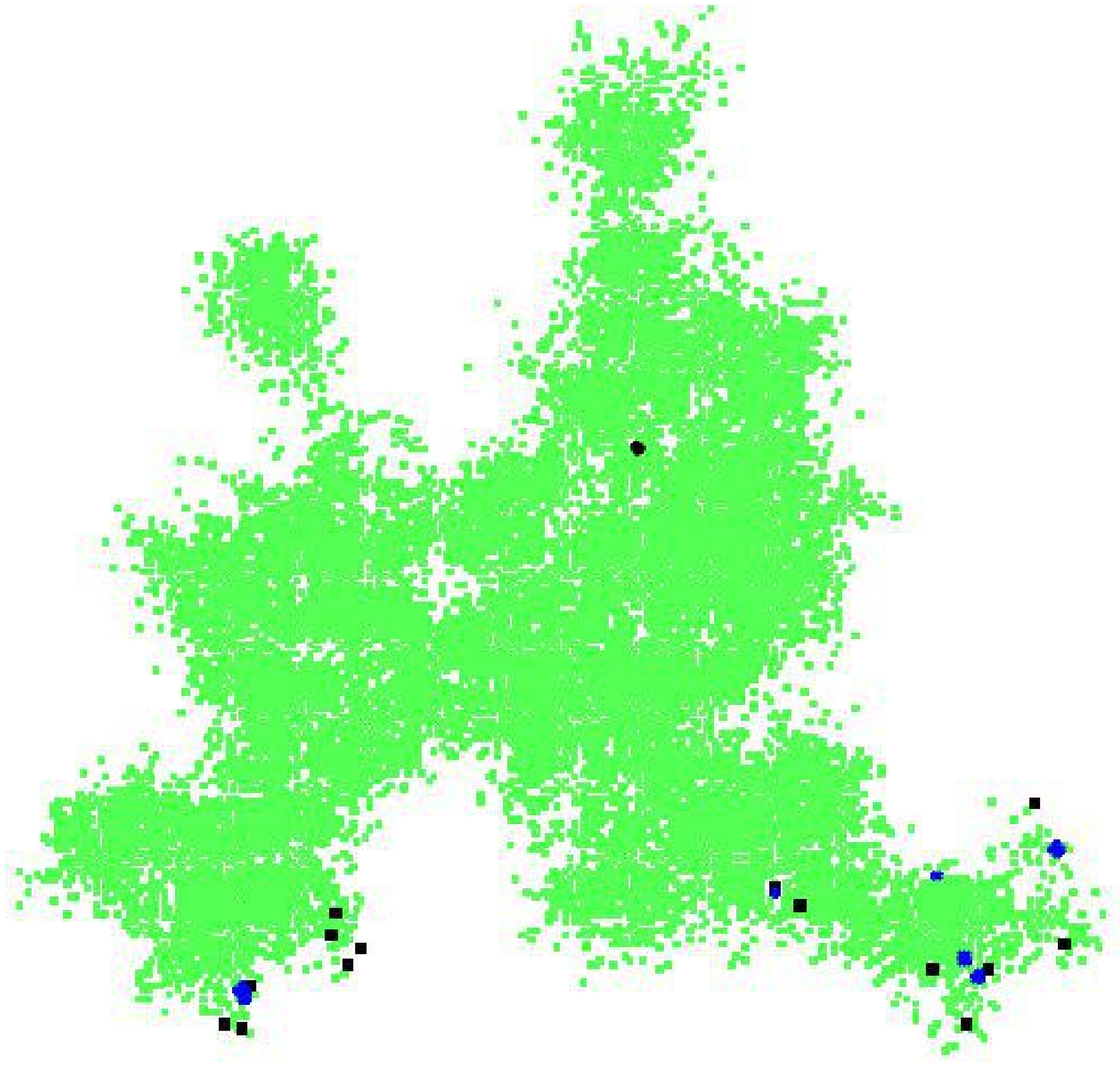}

\caption{\footnotesize{Growing critical clusters at time 2000, for parameters as in Fig.~\ref{wcvda},
with ${\cal D} = 0.02$, $w = w_c = 0.2972$ (left), and ${\cal D} = 0.3$, $w= w_c =0.0812$ (right).
Light color: removed cells;
black dots: transformed cells; black diamond: position of original seed; $\times$:
positions of relatively high signal concentration, $C > 0.1$.
}}
\label{cluster}
\end{center}
\end{figure}

\begin{figure}[h]
\begin{center}
\includegraphics[height=5.5cm,width=7.5cm]{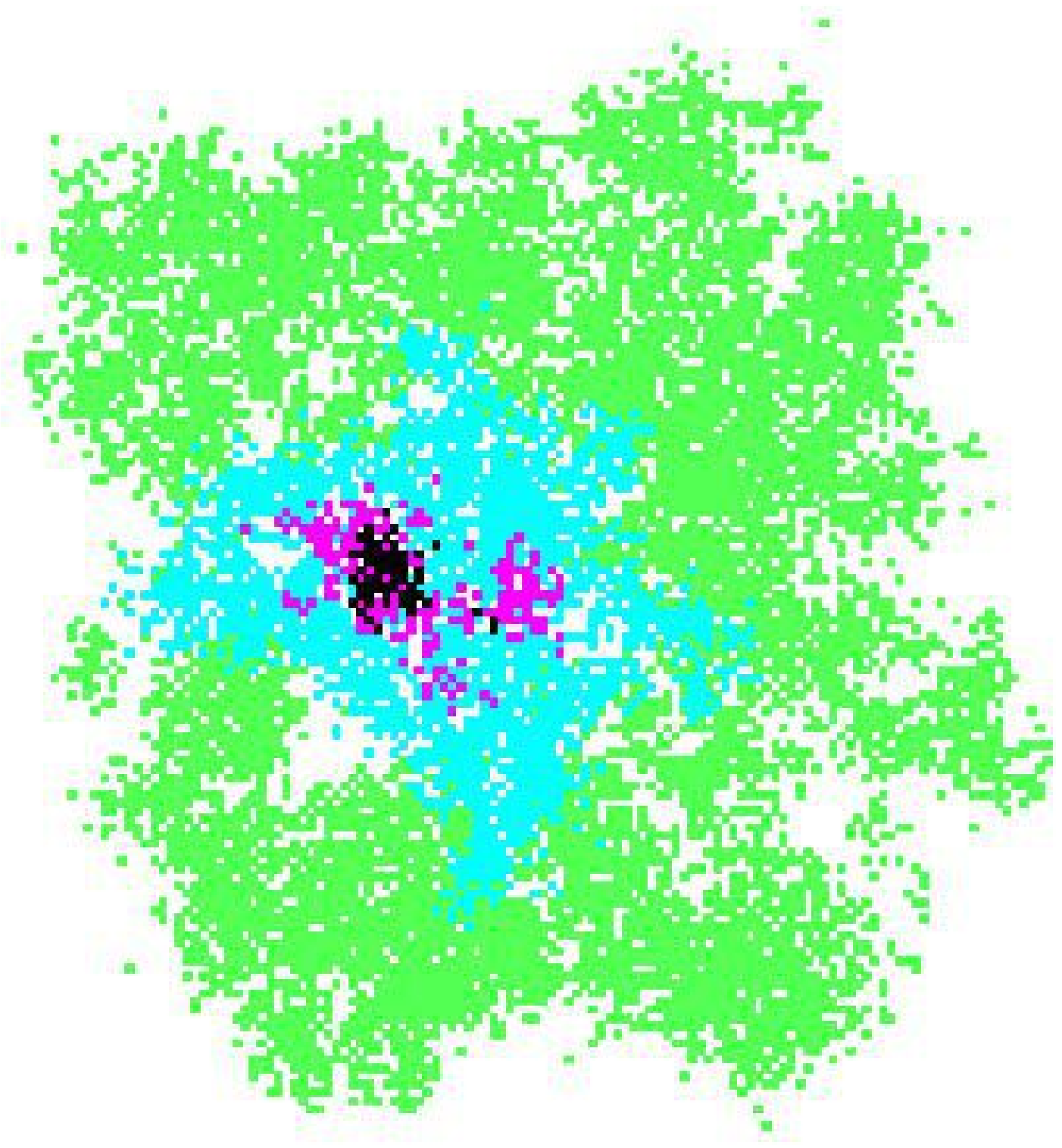}
\includegraphics[height=5.5cm,width=7.5cm]{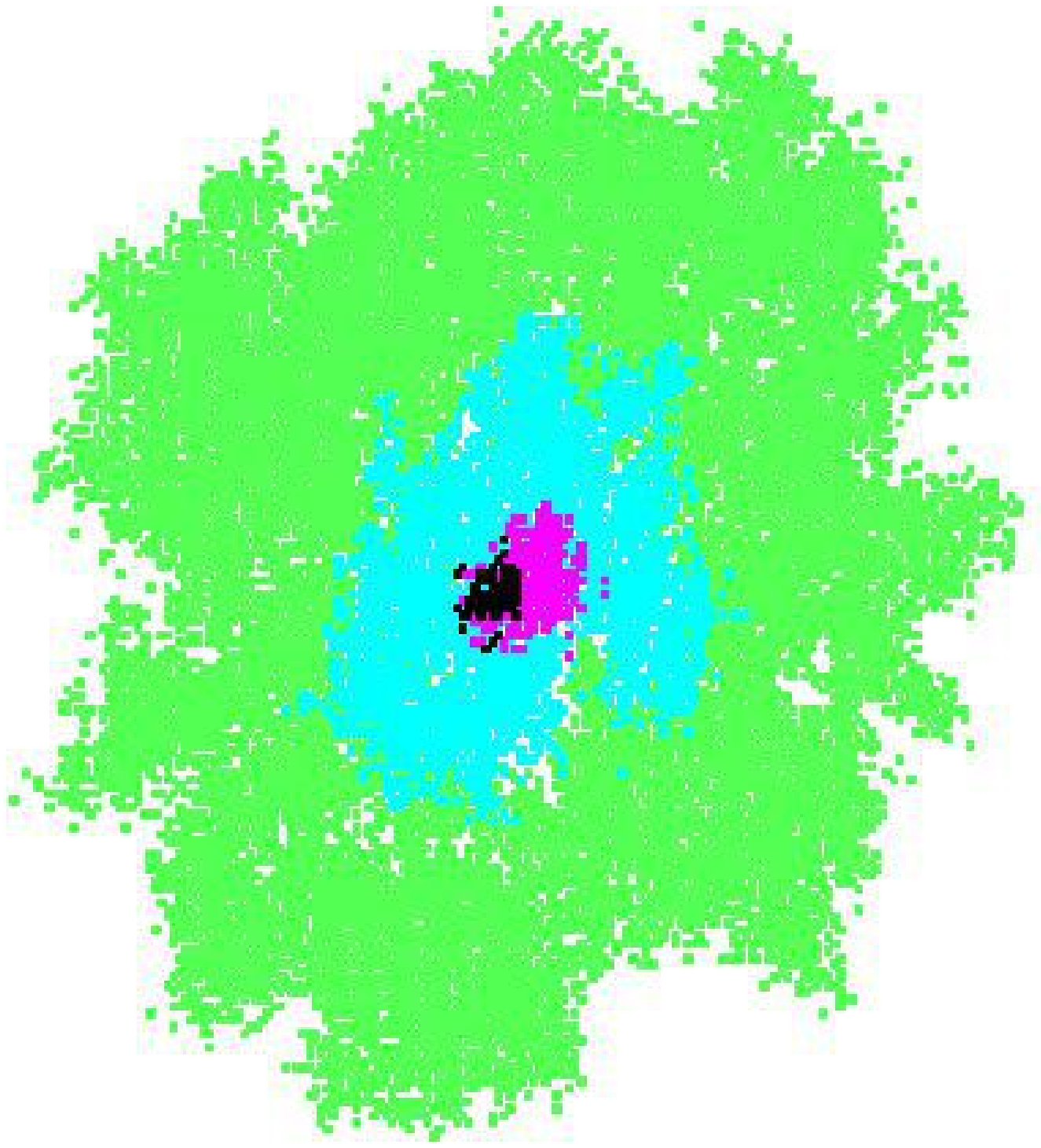}
\vspace{-1cm}

\caption{\footnotesize{Growing clusters in the spreading regime.  Contrasting colors show
the set of removed cells at times 100, 200, 500, and 1000.
Parameters as in Fig.~\ref{wcvda},
with ${\cal D} = 0.1$.  (For these parameters $w_c = 0.1205(5)$.) Left panel: $w = 0.13 \simeq 1.08 w_c $;
right: $w=0.16 \simeq 1.33 w_c$.
}}
\label{growsup}
\end{center}
\end{figure}

Figure \ref{cluster} shows examples of growing clusters at
the critical point for two rather different values of the diffusion rate.
The one corresponding to a smaller ${\cal D}$ value is more densely connected,
while the other shows evidence of ``colonies" growing at some distance from the
main concentration, as well as a more diffuse boundary.
The distribution of transformed and depleted cells around the perimeter is
highly nonuniform in both cases.  Further growth appears to be likely only in
limited areas, as reflected in the nonuniform signal concentration.
In the supercritical regime, cluster growth is more symmetric, but still somewhat irregular,
as shown in Fig.~\ref{growsup}, for parameters such that $w \simeq 1.08 w_c$ and $1.33 w_c$.

\begin{figure}[h]
\begin{center}
\includegraphics[height=7.0cm,width=9.0cm]{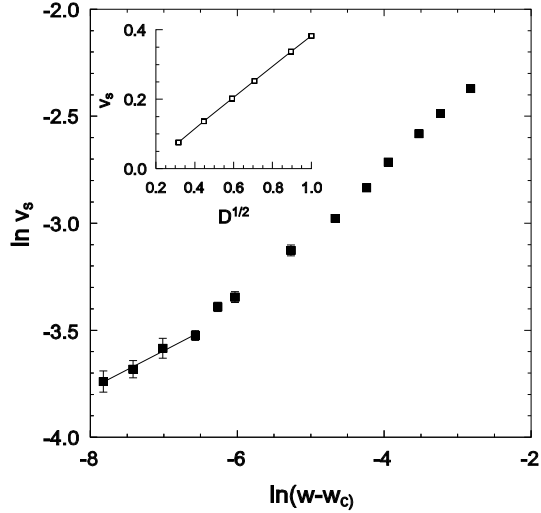}
\caption{\footnotesize{Simulation: spreading velocity $v_s = \langle N_R \rangle^{1/2}/(\sqrt{\pi} t)$
versus $w$ for ${\cal D} = 0.1$.  The slope of the regression line is 0.178.
Inset: spreading velocity versus ${\cal D}^{1/2}$ for $w=0.15$.
Other parameters as in Fig. \ref{ls}.  Lines are a guide for the eye.
}}
\label{vsp}
\end{center}
\end{figure}

We close this section with results on the propagation velocity $v_s$ in the spreading phase.
Near the critical point (or the critical line in the $w$-${\cal D}$ plane) the velocity
is expected to scale as $v_s \sim \epsilon^{\nu_{||} - \nu_\perp}$, where $\epsilon$ is the distance
from criticality \cite{grass1}.  This gives $v_s \sim \epsilon^{0.173}$ for dynamic percolation
in two dimensions.  In simulations, we determine the spreading velocity via the relation
$\langle N_R(t) \rangle \simeq \pi (v_s t)^2$, i.e., the region of removed cells tends, on average,
to a circle of radius $v_s t$ at long times.  In these studies we perform $\sim 100$ realizations
for each $w$ value,
on lattices with $L = 850$ - 1000, extending to a maximum time of $t_M = 8000$ units.
The results (Fig.~\ref{vsp}) are consistent with a power law near the critical point.
A fit to the data, using $w_c = 0.1206$, yields $\nu_{||} - \nu_\perp = 0.178(15)$; including the uncertainty
in $w_c$ itself ($\pm 0.0001$), we obtain $\nu_{||} - \nu_\perp = 0.18(4)$, which, while not
very precise, is consistent with the value expected for dynamic percolation.
The inset of Fig.~\ref{vsp} confirms the
scaling $v_s \sim \sqrt{\cal D}$, as expected on the basis of dimensional analysis,
away from the immediate vicinity of the critical point.

\section{\label{conc} Discussion}

We study an epidemic model consisting of elements (organisms in a community or cells in tissue)
with fixed positions, in which disease or damage is
transmitted by diffusing signals emitted by infected individuals.
The model is formulated on a square lattice in which each site bears
a cell, which can be in one of the states: susceptible,
transformed, depleted, or removed.
Signal diffusion and decay is treated deterministically, given the (random) source distribution
in space and time.
We study the model using mean-field theory (in both simple and diffusive versions)
and Monte Carlo simulation.  Simple MFT predicts the order of magnitude of the critical
value $w_c$, if the diffusion rate is not extremely small, but is insensitive to
the diffusion rate. Diffusive MFT yields a slight
improvement over the simpler analysis; it captures, qualitatively, the fact that $w_c$
decreases with ${\cal D}$, and that $w_c$ diverges as ${\cal D} \to 0$.

The process is found to exhibit a continuous phase transition between spreading and nonspreading
phases.  Simulations of the spread of activity
yield estimates for the critical exponents $\theta$, $\delta$, and $z_{sp}$ consistent
with those of two-dimensional dynamic percolation.
The fractal dimension of the cluster of affected individuals at the critical point is
also consistent with that of dynamic percolation, as are the critical exponents associated with
the subcritical regime, and the scaling of the spreading velocity in the supercritical regime.
Although these results are obtained for a specific set of
parameters, there is little reason to expect a change in scaling behavior for other values,
as long as the diffusion rate is finite.  Indeed, dynamic percolation universality
for epidemic-like processes with a finite range of spreading was already asserted some
time ago \cite{grass1}.  We are unaware, however, of a previous verification of
such behavior in the case of propagation via a diffusing, decaying signal.

The present study suggests several lines for future work.  One is a more detailed
study of the scaling of the spreading velocity.  The ability of mean-field theory or
reaction-diffusion equations to describe this aspect of the process is of interest, as
such approaches are frequently used in applications.
Another subject for future study concerns the nature of the spreading transition
in disordered and fractal media. The possibility of a discontinuous
transition for a nonlinear dependence of the transformation rate on signal concentration
is also worth investigating.  While mean-field theory does predict such a transition
when the concentration-dependent rates are $\propto C^2$, experience with contact-process-like
models suggests that the nature of the transition depends on the details
of the dynamics \cite{grass82,evansCP2}.
Finally, applications to specific processes, such as the radiation-induced
bystander effect, are of great interest, if plausible estimates of the governing rates
can be obtained.

\newpage
\noindent{\bf Acknowledgments}\\

We thank C. H. C. Moreira for helpful discussions.
F. P. Faria is grateful for support provided by Fapemig, Minas Gerais, Brazil.
R. D. acknowledges financial support from CNPq, Brazil.

\bibliographystyle{apsrev}

\end{document}